\documentclass[reprint,showpacs,preprintnumbers,amsmath,amssymb,superscriptaddress,prb,floatfix]{revtex4-2}
\usepackage{graphicx}
\usepackage[utf8]{inputenc}
\usepackage{amssymb}
\usepackage{color}
\usepackage{units}
\usepackage{float}
\usepackage{amsmath}
\usepackage{hyperref} 
\usepackage{epsfig}
\usepackage{bm}
\usepackage{ulem}
\usepackage{hyphenat}
\usepackage{makecell}
\usepackage{color,soul}
\usepackage[dvipsnames]{xcolor}
\usepackage{upgreek}
\usepackage{xcolor}
\usepackage[version=4,arrows=pgf]{mhchem}

\begin{document}
\title{Detecting slow magnetization relaxation via magnetotransport measurements based on the current-reversal method}

\author{Sebastian Beckert}
\affiliation{Institut f{\"u}r Festk{\"o}rper- und Materialphysik, Technische Universit{\"a}t Dresden, 01062 Dresden, Germany}
\author{Richard Schlitz}
\affiliation{Department of Physics, University of Konstanz, 78457 Konstanz, Germany}
\author{Gregor Skobjin}
\affiliation{Department of Physics, University of Konstanz, 78457 Konstanz, Germany}
\author{Antonin Badura}
\affiliation{Institute of Physics ASCR, v.v.i., Cukrovarnick\'a 10, 162 53, Prague, Czech Republic}
\author{Miina Leiviskä}
\affiliation{Institute of Physics ASCR, v.v.i., Cukrovarnick\'a 10, 162 53, Prague, Czech Republic}
\affiliation{Univ. Grenoble Alpes, CNRS, CEA, Grenoble INP, IRIG-Spintec, F-38000 Grenoble, France}
\author{Dominik Kriegner}
\affiliation{Institut f{\"u}r Festk{\"o}rper- und Materialphysik, Technische Universit{\"a}t Dresden, 01062 Dresden, Germany}
\affiliation{Institute of Physics ASCR, v.v.i., Cukrovarnick\'a 10, 162 53, Prague, Czech Republic}
\author{Daniel Scheffler}
\affiliation{Institut f{\"u}r Festk{\"o}rper- und Materialphysik, Technische Universit{\"a}t Dresden, 01062 Dresden, Germany}
\affiliation{Institute of Physics ASCR, v.v.i., Cukrovarnick\'a 10, 162 53, Prague, Czech Republic}
\author{Giacomo Sala}
\affiliation{Department of Materials, ETH Zurich, CH-8093 Zurich, Switzerland}
\author{Kamil Olejník}
\affiliation{Institute of Physics ASCR, v.v.i., Cukrovarnick\'a 10, 162 53, Prague, Czech Republic}
\author{Lisa Michez}
\affiliation{Aix-Marseille Univ, CNRS, CINaM, Marseille, France}
\author{Vincent Baltz}
\affiliation{Univ. Grenoble Alpes, CNRS, CEA, Grenoble INP, IRIG-Spintec, F-38000 Grenoble, France}
\author{Andy Thomas}
\affiliation{Institut f{\"u}r Festk{\"o}rper- und Materialphysik, Technische Universit{\"a}t Dresden, 01062 Dresden, Germany}
\affiliation{Leibniz Institute of Solid State and Materials Science (IFW Dresden), Helmholtzstr.\ 20, 01069 Dresden, Germany}
\author{Helena Reichlová}
\affiliation{Institut f{\"u}r Festk{\"o}rper- und Materialphysik, Technische Universit{\"a}t Dresden, 01062 Dresden, Germany}
\affiliation{Institute of Physics ASCR, v.v.i., Cukrovarnick\'a 10, 162 53, Prague, Czech Republic}
\author{Sebastian T. B. Goennenwein}
\affiliation{Department of Physics, University of Konstanz, 78457 Konstanz, Germany}

\date{May 23, 2024}

\begin{abstract}
Slow magnetization relaxation processes are an important time-dependent property of many magnetic materials. We show that magnetotransport measurements based on a well-established current-reversal  method can be utilized to implement a simple and robust screening scheme for such relaxation processes. We demonstrate our approach considering the anomalous Hall effect in  a Pt/Co/AlO$_x$ trilayer model system, and then explore relaxation in $\tau$-MnAl films. Compared to magnetotransport experiments based on ac lock-in techniques, we find that the dc current-reversal method is particulary sensitive to relaxation processes with relaxation time scales on the order of seconds, comparable to the current-reversal measurement time scales.
\end{abstract}
\maketitle
\section{Introduction}
\label{sec:intro}
Slow magnetization relaxation processes in magnetic materials, i.e., the delayed response of the magnetization to the external magnetic field \cite{Preisach1935},  have been observed and investigated for several decades, as described by Néel \cite{Neel1951}, Preisach \cite{Preisach1935}, and others \cite{Wohlfarth1984,Bayreuther1989,Labrune1989,Pommier1990,Thiele1995,Fry1999,Fry2000,Xi2008,Xie2008}. Over time, slow relaxation processes have been reported in materials with various magnetic properties, including thin films with perpendicular magnetic anisotropy (PMA), spin glasses, and compensated magnets \cite{Labrune1989,Ferre2002,Xu2022,Kaspar2021,Xi2008,Pasco2021}. Magnetic relaxation can be probed by the magneto-optical Kerr effect (MOKE) \cite{Thiele1995,Fry1999,Fry2000,Labrune1989}, magnetometry \cite{Bayreuther1989,Pasco2021}, magnetic microscopy \cite{Weir1999,Pommier1990}, and the anomalous Hall effect (AHE) \cite{Xie2008,Bhatt2022,Kan2020}. The relaxation processes also impact magnetic characterization experiments. For example, the coercive field measured in recording tape depends on the field sweep rate owing to relaxation effects \cite{Flanders1987}. Similar observations have also been reported in AHE experiments on thin films \cite{Kan2020}. Indeed, the time necessary for the reversal of the magnetization to reach a new (quasi-) equilibrium state can vary over many orders of magnitude, depending on various parameters such as the material, sample size, defects, temperature, and external magnetic field strength. It can be quasi-instantaneous \cite{Wongsam1999}, or reach up to several hundreds or thousands of seconds \cite{Labrune1989,Bayreuther1989}. In the following, we limit the considerations to slow magnetization relaxation, i.e., relaxation processes which happen on time scales ranging from a few seconds \cite{Thiele1995,Labrune1989,Pommier1990} up to thousands of seconds \cite{Labrune1989}.   \\ 
In-depth studies of slow magnetization relaxation processes require either the measurement of the time evolution of the magnetization over long time intervals,  at fixed magnetic field, or a series of magnetic field sweeps with a systematic variation in field sweep rates \cite{Kan2020}. Since such experiments can be very time consuming, slow relaxation is often ignored or goes unnoticed. We here show that a current-reversal method widely used for sensitive magnetotransport and AHE measurements \cite{Asa2020,Rueffer2009} allows to straightforwardly screen for the presence of slow magnetization relaxation processes. The screening benefits from the slow time-scale of the current-reversal measurement as compared to conventional lock-in measurements, and provides quantitative access to the normalized relaxation parameter \(\frac{\mathrm{d}M_\mathrm{norm}}{\mathrm{d}t}\) (with the normalized magnetization \(M_\mathrm{norm}=\frac{M(H)}{M_\mathrm{s}}\) and the saturation magnetization $M_\mathrm{s}$). In turn, if unaccounted for, relaxation effects can result in measurement artifacts.\\
This paper is structured in the following way: In Sec. \ref{sec:curr_rev} we introduce the current-reversal method and derive the effects of slow magnetic relaxation on this method. In Sec. \ref{sec:DC_PtCo} we apply the current-reversal method to a Pt/Co/AlO$_x$ trilayer showing slow magnetic relaxation to corroborate the considerations from Sec. \ref{sec:curr_rev}. We then compare the dc current-reversal method to the ac lock-in method in Sec. \ref{sec:AC_PtCo}, and finally apply the method to $\tau$-MnAl as a material of current interested in Sec. \ref{sec:DC_MnAl}. 
\section{Current-reversal method}
\label{sec:curr_rev}
The current-reversal method (in the following also called delta method) is a standard technique to reduce or even eliminate thermoelectric voltages from electrical measurements \cite{Pitsakis1989,Rueffer2009,Goennenwein2015}. In this current-bias, four-point measurement method, for each measurement point, e.g. for each magnetic field magnitude in a magnetic field sweep, two data points are recorded: one voltage $V(+I)$ for positive current polarity and one voltage $V(-I)$  for negative current polarity. The delta minus signal $V_{\Delta-}$ and delta plus signal $V_{\Delta+}$ are then calculated as:
\begin{equation}
    V_{\Delta-} = \frac{V(+I)-V(-I)}{2}
    \label{eq:simple_delta_minus}
\end{equation}
and
\begin{equation}
    V_{\Delta+} = \frac{V(+I)+V(-I)}{2}
    \label{eq:simple_delta_plus}
\end{equation}
respectively.
Consequently,  $V_{\Delta-}$ contains all effects odd in current, such as the resistive contribution. For many experiments focused on (magneto-)resistance, this is the only quantity of interest. In turn, $V_{\Delta+}$ contains all effects even in current, in particular the above mentioned thermoelectric effects. While thermoelectric contributions are often neglected in transport measurements, $V_{\Delta+}$ has proven very useful in the study of, e.g., the spin Seebeck effect \cite{Schreier2013,Gueckelhorn2020,Cramer2019}, or spin torque effects \cite{Manchon2019,Garello2013}. In the following, we demonstrate that the $V_{\Delta+}$ signal can also be used to screen for slow relaxation processes.\\
In a first step, we discuss the consequences of slow magnetic relaxation effects on current-reversal Hall measurements. Phenomenologically, the Hall voltage $V_\mathrm{H}$ in a ferromagnet with Hall bar or Hall cross geometry is described by \cite{Nagaosa2010,Pugh1932,Pugh1953}:
\begin{equation}
    V_\mathrm{H} = \frac{I}{d}\cdot A_\mathrm{OHE}\cdot\mu_0 H_\mathrm{ext} + \frac{I}{d}\cdot A_\mathrm{AHE}\cdot\mu_0 M(H_\mathrm{ext})
    \label{eq:Hall}
\end{equation}
with the sourced current $I$, the ordinary and anomalous Hall coefficients $A_\mathrm{OHE}$ and $A_\mathrm{AHE}$, the external magnetic field $H_\mathrm{ext}$, and the sample thickness $d$. In Hall voltage measurements, voltage offsets due to a geometrical misalignment of the voltage probes or similar artifacts are commonplace \cite{Badura_Arxiv,DeGrave2013}. This offset is often subtracted from the data without further consideration. Moreover, also a planar Hall effect (PHE)/ transversal anisotropic magnetoresistance contribution could be present in the measured  transversal `Hall' voltage. This PHE contribution would depend on the magnetization orientation, but in contrast to the AHE it would be even under the reversal of magnetic field. For simplicity, we assume that the AHE is much larger than the PHE in the following, which is reasonable for a ferromagnet with high PMA, such that the latter can be neglected \cite{Ky1968,Garello2013}.\\
To formally model the role of slow magnetic relaxation on the current-reversal method, the time-dependent relaxation of the magnetization $M(t)$ must be taken to account. Owing to the relaxation, the magnetization at times $t$ and $t+\Delta t$ can differ, as approximated by a first order Taylor expansion:
\begin{equation}
    M(t+\Delta t) = M(t)+\frac{\mathrm{d} M}{\mathrm{d}t}\cdot\Delta t + \mathcal{O}(\Delta t^2)
    \label{eq:Magnetization}
\end{equation}
Here the derivative \(\frac{\mathrm{d}M}{\mathrm{d}t}\), which we will call relaxation parameter in the following, describes the relaxation. Using Eqs. \ref{eq:simple_delta_plus} and \ref{eq:Hall}, the voltage in the delta plus signal becomes:
\begin{equation}
    V_{\mathrm{H},\Delta+}\left(t+\frac{\Delta t}{2}\right) = \frac{V_\mathrm{H}(t,+I)+V_\mathrm{H}(t+\Delta t,-I)}{2}
    \label{eq:V_delta_plus_pre}
\end{equation}
Note that we use \(t+\frac{\Delta t}{2}\) as time point for the delta plus voltage $V_{\mathrm{H},\Delta +}$ since the value is calculated from voltages measured at the time points $t$ and $t+\Delta t$ with $\Delta t$, being the time between the two voltage readings. Using Eq. \ref{eq:Hall}, Eq. \ref{eq:Magnetization} and Eq. \ref{eq:V_delta_plus_pre}, we can therefore write:
\begin{equation}
    \begin{split}
    V_{\mathrm{H},\Delta+}\left(t+\frac{\Delta t}{2}\right) &=  - \frac{I}{2d} A_\mathrm{AHE}\mu_0\frac{\mathrm{d}M}{\mathrm{d}t}\bigg|_{t+\frac{\Delta t}{2}}\Delta t \\\ &= - \frac{1}{2}\frac{\mathrm{d}V_\mathrm{H}}{\mathrm{d}t}\bigg|_{t+\frac{\Delta t}{2}}\Delta t
    \end{split}
    \label{eq:V_delta_plus}
\end{equation}
We would like to stress, that the negative sign in Eq. \ref{eq:V_delta_plus} comes from the fact that the first measurement point (at time $t$) is taken with a positive current polarity, while the second measurement point (at time $t + \Delta t$) is taken for a negative current polarity. Instead of the voltage, one can also consider the resistance \(R_{\mathrm{H},\Delta+}=\frac{V_{\mathrm{H},\Delta+}}{|I|}\) calculated from this voltage, which then is:
\begin{equation}
\begin{split}
    R_{\mathrm{H},\Delta+}\left(t+\frac{\Delta t}{2}\right) &=  - \frac{1}{2d} A_\mathrm{AHE}\mu_0\frac{\mathrm{d}M}{\mathrm{d}t}\bigg|_{t+\frac{\Delta t}{2}}\Delta t \\\ &= - \frac{1}{2}\frac{\mathrm{d}R_\mathrm{H}}{\mathrm{d}t}\bigg|_{t+\frac{\Delta t}{2}}\Delta t
\end{split}
\label{eq:R_trans_plus}
\end{equation}
Eqs. \ref{eq:V_delta_plus} and \ref{eq:R_trans_plus} imply that the delta plus signal allows to quantify the relaxation parameter $\frac{\mathrm{d}M}{\mathrm{d}t}$ (or rather the normalized relaxation parameter $\frac{\mathrm{d}M_\mathrm{norm}}{\mathrm{d}t}$), i.e.,  the presence and amplitude of slow relaxation in a magnetic conductor, given that the time scale of the relaxation is sufficiently large compared to the time step $\Delta t$, and that the signal $\frac{\mathrm{d}V_\mathrm{H}}{\mathrm{d}t}\Delta t$ is above the noise floor of the voltage measurement. It is important to stress that the time it takes to acquire a single voltage reading can be fast. In the discussion above, we thus have tacitly assumed that the voltage reading is instantaneous, while a finite waiting time $\Delta t$ separates the two measurement points with opposite current polarity. In current-reversal measurements, typical voltage acquisition times range from \unit[0.2]{s} to \unit[1]{s}, while the delay $\Delta t$ is of the order of several seconds (see Sec. \ref{sec:DC_PtCo} and \ref{sec:DC_MnAl}). The voltage acquisition thus is fast compared to $\Delta t$. If the time required to acquire one voltage reading becomes comparable to $\Delta t$, relaxation also will impact the voltage reading itself, and the simple Eqs. \ref{eq:V_delta_plus}-\ref{eq:R_delta_minus} are no longer applicable. However, qualitatively, the delta plus signal will still show the characteristic signatures of relaxation.\\
Another important question is, which quantitative information can be extracted from the delta plus signal in the Hall resistance. Considering  the Hall voltage given by Eq. \ref{eq:Hall}, the anomalous Hall signal is proportional to the magnetization, and the OHE proportional to the external magnetic field. Therefore it is possible to determine the OHE contribution by fitting a linear function to the Hall voltages at high magnetic fields (i.e., in magnetic saturation) and to remove the OHE by subtraction. Since the AHE contribution is proportional to the (out-of-plane) magnetization component of the sample, we can therefore calculate a normalized AHE signal, which is equivalent to the normalized magnetization $M_\mathrm{norm}$ \cite{Lindemuth2004}:
\begin{equation}
    M_\mathrm{norm}(H) = \frac{M(H)}{M_\mathrm{s}}=\frac{R_\mathrm{AHE}(H)}{R_\mathrm{AHE,s}}
    \label{eq:M_norm}
\end{equation}
with the field dependent magnetization $M(H)$, the saturation magnetization $M_\mathrm{s}$, the field dependent anomalous Hall resistance $R_\mathrm{AHE}(H)$ and the anomalous Hall resistance at saturation $R_\mathrm{AHE,s}$. In this way, the transport experiments yield: \(\frac{\mathrm{d}M_\mathrm{norm}}{\mathrm{d}t}\) as a quantitative measure of relaxation. Using Eq. \ref{eq:M_norm} and the expression for $R_{\mathrm{H},\Delta +}$ (see Eq. \ref{eq:R_trans_plus}):
\begin{equation}
    R_{\mathrm{H},\Delta +} = - \frac{1}{2}\frac{\mathrm{d}R_\mathrm{H}}{\mathrm{d}t}\Delta t
\end{equation}
we obtain:
\begin{equation}
    \frac{\mathrm{d}M_\mathrm{norm}}{\mathrm{d}t} = \frac{\mathrm{d}R_\mathrm{AHE}}{\mathrm{d}t}\frac{1}{R_{\mathrm{AHE,s}}} = - 2 \frac{R_ {\mathrm{H},\Delta+}}{R_\mathrm{AHE,s}\Delta t}
    \label{eq:time_deriv}
\end{equation}
Note that we hereby have assumed that \(\frac{\mathrm{d}{R_\mathrm{AHE}}}{\mathrm{d}t}=\frac{\mathrm{d}R_{\mathrm{H}}}{\mathrm{d}t}\). This is reasonable, since the OHE is time-independent (instantaneous on the time scales considered here) and therefore the AHE is the only time-dependent contribution in the Hall resistance.
This means we can directly calculate the normalized relaxation parameter \(\frac{\mathrm{d}M_\mathrm{norm}}{\mathrm{d}t}\) (at the beginning of the relaxation) from the delta plus signal of the Hall resistance. With knowledge of the saturation magnetization it is further possible to estimate the relaxation parameter \(\frac{\mathrm{d}M}{\mathrm{d}t}\). Moreover, using the normalized magnetization from Eq. \ref{eq:M_norm} to calculate a normalized magnetic susceptibility \(\frac{\mathrm{d}M_\mathrm{norm}}{\mathrm{d}\mu_0 H}\), we estimate the change of the measured coercive field, if the magentic field sweep rate is changed:
\begin{equation}
    \frac{\mathrm{d}\mu_0H_\mathrm{c}}{\mathrm{d}t} = \frac{\frac{\mathrm{d}M_\mathrm{norm}}{\mathrm{d}t}}{\frac{\mathrm{d}M_\mathrm{norm}}{\mathrm{d}\mu_0H}} = \frac{\mathrm{d}M_\mathrm{norm}}{\mathrm{d}t} \cdot \frac{\mathrm{d}\mu_0H}{\mathrm{d}M_\mathrm{norm}}
    \label{eq:dHdt}
\end{equation}
By analogy to the delta plus signal, the voltage for the delta minus signal is:
\begin{equation}
    \begin{split}
    V_{\mathrm{H},\Delta-}\left(t+\frac{\Delta t}{2}\right)=\frac{I}{d}\bigg(&A_\mathrm{OHE}\mu_0H_\mathrm{ext}+A_\mathrm{AHE}\mu_0M(t)\\\ &+\frac{1}{2}A_\mathrm{AHE}\mu_0\frac{\mathrm{d}M}{\mathrm{d}t}\bigg|_{t+\frac{\Delta t}{2}}\Delta t\bigg)
    \end{split}
\end{equation}
Thus:
\begin{equation}
    V_{\mathrm{H},\Delta-}\left(t+\frac{\Delta t}{2}\right) = V_\mathrm{H}\left(t\right) + \frac{1}{2}\frac{\mathrm{d}V_\mathrm{H}}{\mathrm{d}t}\bigg|_{t+\frac{\Delta t}{2}}\Delta t
\end{equation}
and \(R_{\mathrm{H},\Delta-} = \frac{V_{\mathrm{H},\Delta-}}{|I|}\):
\begin{equation}
    R_{\mathrm{H},\Delta-}\left(t+\frac{\Delta t}{2}\right) = R_\mathrm{H}\left(t\right) + \frac{1}{2}\frac{\mathrm{d}R_\mathrm{H}}{\mathrm{d}t}\bigg|_{t+\frac{\Delta t}{2}}\Delta t
    \label{eq:R_delta_minus}
\end{equation}
Slow relaxation (finite $\frac{\mathrm{d}M}{\mathrm{d}t}$) thus also results in a finite contribution to the delta minus signal of the Hall resistance. However this contribution is much smaller than the (time-independent) part of the Hall effect in most materials, and thus difficult to resolve.\\
Finally, substantial relaxation effects typically are observed when the magnetic susceptibility $\frac{\mathrm{d}M}{\mathrm{d}H}$ is large. In a more elaborate model, also the relaxation time constant can depend on the magnetic field strength \cite{Labrune1989}. We thus expect marked relaxation-related contributions to the delta plus signal of the Hall resistance $R_{\mathrm{H},\Delta +}$ in the field range for magnetic hysteresis, in which the magnetization (and therefore AHE) is switching.
\section{Magnetic relaxation in a  P\lowercase{t}/C\lowercase{o}/A\lowercase{l}O$_x$ thin film heterostructure}
\label{sec:DC_PtCo}
\begin{figure}
	\centering
	\includegraphics{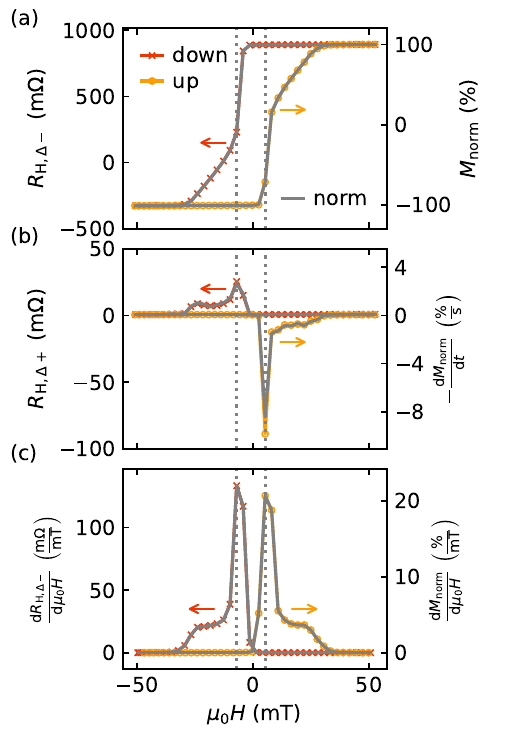}
	\caption{\textbf{Hall resistance of  Pt/Co/AlO$_x$, recorded using the delta method:} \\(a) Left axis: delta minus signal of the Hall resistance, right axis: normalized magnetization calculated from the Hall resistance using Eq. \ref{eq:M_norm}.\\(b) Left axis: delta plus signal of the Hall resistance, right axis: negative of (normalized) relaxation parameter \(\frac{\mathrm{d}M_\mathrm{norm}}{\mathrm{d}t}\) calculated from the delta plus signal of the Hall resistance, using Eq. \ref{eq:time_deriv}. \\(c) Left axis: derivative of the delta minus signal of the Hall resistance with respect to the external field $\mu_0 H$, right axis: normalized susceptibility (derivative \(\frac{\mathrm{d}M_\mathrm{norm}}{\mathrm{d}\mu_0H}\) calculated from the normalized magnetization). The data was recorded with a time step $\Delta t=\unit[3]{s}$ and $I=\unit[1]{mA}$ at room temperature. The field sweep directions are indicated by arrows and different colors. \label{fig:Delta_plus_delta_minus_field_sweep_Co_Pt}}
\end{figure}

In order to experimentally corroborate the consideration in Sec. \ref{sec:curr_rev}, we study the magnetotransport response of a trilayer consisting of \unit[5]{nm} platinum, \unit[1.1]{nm} cobalt and \unit[1.6]{nm} aluminum oxide, sputter-deposited onto a silicon substrate with thermal silicon oxide. We refer to this trilayer as Pt/Co/AlO$_x$. The heterostructure is patterned into Hall crosses, as described in more detail by Nabben \textit{et al.} \cite{Nabben_Arxiv}. We measure the Hall effect using a room-temperature electromagnet setup, with the magnetic field applied along the film normal. The current is hereby applied using a Keithley 2450 sourcemeter and the transversal (Hall) voltage is recorded by a Keithley 2182 nanovoltmeter.\\  The sample shows perpendicular magnetic anisotropy, a high remanence, and a deviation from a square hysteresis loop at higher magnetic fields, as evident in the Hall curve in Fig. \ref{fig:Delta_plus_delta_minus_field_sweep_Co_Pt} (a). The delta minus signal of the Hall resistance \(R_{\mathrm{H},\Delta-}=\frac{V_{\mathrm{H},\Delta-}}{|I|}\) is dominated by the anomalous Hall contribution \cite{Nagaosa2010}, the ordinary Hall contribution is negligible in comparison (see left axis of Fig. \ref{fig:Delta_plus_delta_minus_field_sweep_Co_Pt}(a)). Similar hysteresis curves were observed in other Co/Pt layers, as for example described by Shen \textit{et al.} for Co/Pt multilayers consisting out of 30 stacks. These authors also attribute the shape of the hysteresis loop to non-uniform domain wall expansion \cite{Shen1993}. We expect similar effects in our Pt/Co/AlO$_x$ film. We intentionally did not remove the offset in the Hall measurements discussed above in Sec. \ref{sec:curr_rev}. The right axis of Fig. \ref{fig:Delta_plus_delta_minus_field_sweep_Co_Pt}(a) shows the normalized magnetization of the sample calculated with Eq. \ref{eq:M_norm}.\\
% 20231030 nk220210c2 cross2 RT 1mA
We now consider the delta plus signal of the Hall resistance \(R_{\mathrm{H},\Delta+}=\frac{V_{\mathrm{H},\Delta+}}{|I|}\) shown in Fig. \ref{fig:Delta_plus_delta_minus_field_sweep_Co_Pt}(b) (left axis), obtained from the same set of data as $R_{\mathrm{H},\Delta-}$, simply by adding instead of subtracting subsequent voltage measurements. As expected from the discussion in Sec.  \ref{sec:curr_rev} we indeed observe a finite signal. We further plot $\frac{\mathrm{d}R_{\mathrm{H},\Delta-}}{\mathrm{d}\mu_0 H}$ as a measure for the magnetic susceptibility $\frac{\mathrm{d}M}{\mathrm{d}H}$ in the left axis of Fig. \ref{fig:Delta_plus_delta_minus_field_sweep_Co_Pt}(c), and obtain sharp peak followed by a broad shoulder in the magnetic field ranges in which the magnetization reorients, which is qualitatively similar to the delta plus signal $R_{\mathrm{H},\Delta+}$ in Fig. \ref{fig:Delta_plus_delta_minus_field_sweep_Co_Pt}(b). The magnetic field sweep direction is hereby encoded in the sign of $R_{\mathrm{H},\Delta+}$, while $\frac{\mathrm{d}R_{\mathrm{H},\Delta-}}{\mathrm{d}\mu_0 H}$ is not affected by the sweep direction, since the time axis is irrelevant here. Moreover, we have carefully checked that the delta plus signal shows an ohmic behavior, i.e., delta plus signal of the Hall resistance $R_{\mathrm{H},\Delta+}$ is constant for different magnitudes of the sourced current $I$. This together with the signal's shape rules out the possibility of a thermal origin. Nevertheless, similar effects could be misinterpreted as thermovoltages, if the delta method is used to intentionally measure thermovoltages.
In Figs. \ref{fig:Delta_plus_delta_minus_field_sweep_Co_Pt}(b) and \ref{fig:Delta_plus_delta_minus_field_sweep_Co_Pt}(c), the position and magnitude of the peaks in Fig. \ref{fig:Delta_plus_delta_minus_field_sweep_Co_Pt}(b) and (c) are not symmetrical. This is due to the fact that in our setup, we apply a series of current levels to the coils of the electromagnet to generate different magnetic field values. Owing to the finite remanence of the iron yoke of the magnet, this results in  slightly different applied magnetic fields for positive and negative currents in the coils. Since the peaks are very narrow, small deviations in the magnetic field can have a big influence on the measured value.
Additionally, the reason for the qualitative difference of the two-step variation with $H$ in Fig. \ref{fig:Delta_plus_delta_minus_field_sweep_Co_Pt}(b) and (c) could be a field-dependent time constant of the relaxation.\\
The negative of the normalized relaxation parameter \(\frac{\mathrm{d}M_\mathrm{norm}}{\mathrm{d}t}\) (see Eq. \ref{eq:time_deriv}) is shown in the right axis of Fig. \ref{fig:Delta_plus_delta_minus_field_sweep_Co_Pt}(b). It goes up to about \unitfrac[8]{\%}{s} at the sharp peak and is in the order of \unitfrac[0.8]{\%}{s} in the region of the shoulders. In the right axis of Fig. \ref{fig:Delta_plus_delta_minus_field_sweep_Co_Pt}(c) we plot the normalized susceptibility \(\frac{\mathrm{d}M_\mathrm{norm}}{\mathrm{d}\mu_0H}\). With the approximation from Eq. \ref{eq:dHdt}, the normalized relaxation parameter \(\frac{\mathrm{d}M_\mathrm{norm}}{\mathrm{d}t}\) and the normalized susceptibility \(\frac{\mathrm{d}M_\mathrm{norm}}{\mathrm{d}\mu_0H}\) (see right axis of Fig. \ref{fig:Delta_plus_delta_minus_field_sweep_Co_Pt}(c)) we can therefore estimate the sensitivity of the measured coercive field with regard to small changes of the field sweep rate or measurement delays as up to approx. \unitfrac[0.4]{mT}{s}, for a coercive field of approx. \unit[7]{mT}. Here the derivative \(\frac{\mathrm{d}H_\mathrm{c}}{\mathrm{d}t}\) is in the same order of magnitude, as the coercive field. Therefore changes of the field sweep rate will have a noticeable impact on the measurement of the coercive field.\\
\begin{figure}
	\centering
	\includegraphics{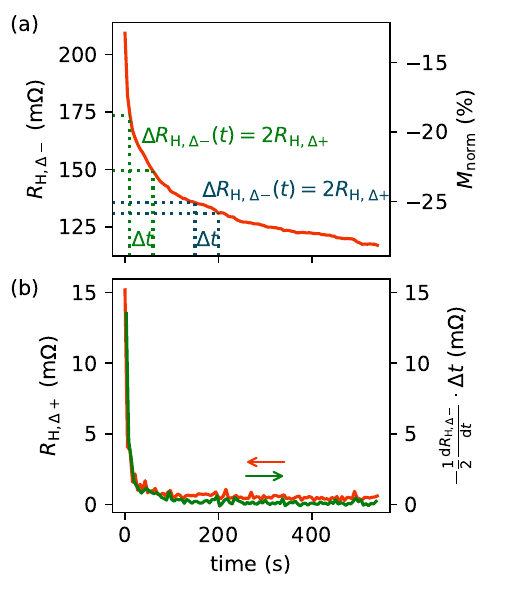}
	\caption{\textbf{Time-dependent evolution of the Hall resistance in Pt/Co/AlO$_x$ film at $\mu_0 H = \unit[-7]{mT}$:}\\
	(a) Left axis: measured time trace of the delta minus signal of the Hall resistance, with schematic depiction of the link between $R_{\mathrm{H},\Delta-}$ and $R_{\mathrm{H},\Delta+}$ in the delta method with exaggerated time steps $\Delta t$, right axis: time trace of normalized magnetization calculated from $R_{\mathrm{H},\Delta-}$ using Eq. \ref{eq:M_norm}. \\(b) Left axis: measured time trace of the delta plus signal of the Hall resistance (red) and right axis: derivative $-\frac{1}{2}\frac{\mathrm{d}R_{\mathrm{H},\Delta-}}{\mathrm{d}t}$ (green) calculated via Eq. \ref{eq:R_trans_plus}.
	\label{fig:Decay_sampling_points}}
\end{figure}
% 20231030 Pt/Co/AlOx nk220210c2 cross2 RT 1mA -7mT = -0.8A\\
To confirm that the delta plus signal of the Hall resistance originates from slow magnetic relaxation, we measure a time trace of both the delta plus and delta minus signal of the Hall resistance. In these measurements, we first saturate the magnetization of the thin film  by applying a magnetic field of \unit[50]{mT}. Then we sweep the magnetic field to a given magnetic field value (i.e., $\unit[-7]{mT}$ in Fig. \ref{fig:Decay_sampling_points}). Now the magnetic field is kept constant and the voltage $V(t)$ is recorded over time, using the delta method. The investigated Pt/Co/AlO$_x$ films indeed exhibit a slow magnetic relaxation as shown by the time trace of the delta minus signal of the Hall resistance $R_{\mathrm{H},\Delta-}$ in Fig. \ref{fig:Decay_sampling_points}(a). In the time trace of the delta plus signal of the Hall resistance $R_{\mathrm{H},\Delta+}$, depicted in Fig. \ref{fig:Decay_sampling_points}(b) (red curve), we observe a finite signal which decays over time. This decay is equivalent to the quantity $- \frac{1}{2}\frac{\mathrm{d}R_\mathrm{H}}{\mathrm{d}t}\Delta t$ calculated in Eq. \ref{eq:R_trans_plus} and can be seen by comparing the red and green curve in Fig. \ref{fig:Decay_sampling_points}(b). 
Note that in Eq. \ref{eq:R_trans_plus} we have the derivative $\frac{\mathrm{d}R_{\mathrm{H}}}{\mathrm{d}t}$, while in Fig. \ref{fig:Decay_sampling_points}(b) the derivative $\frac{\mathrm{d}R_{\mathrm{H},\Delta -}}{\mathrm{d}t}$ is plotted. Strictly speaking, these two terms are not equivalent. We can write the time derivative of Eq. \ref{eq:R_delta_minus} as:
\begin{equation}
    \frac{\mathrm{d}R_{\mathrm{H},\Delta -}}{\mathrm{d}t} = \frac{\mathrm{d}R_{\mathrm{H}}}{\mathrm{d}t} + \frac{1}{2} \frac{\mathrm{d}^2R_{\mathrm{H},\Delta -}}{\mathrm{d}t^2}\Delta t
    \label{eq:dRdt}
\end{equation}
If we transpose Eq. \ref{eq:dRdt}  to $\frac{\mathrm{d}R_{\mathrm{H}}}{\mathrm{d}t}$ and use this term in Eq. \ref{eq:R_trans_plus} we obtain:
\begin{equation}
\begin{split}
    R_{\mathrm{H},\Delta+}\left(t+\frac{\Delta t}{2}\right) =&  - \frac{1}{2}\frac{\mathrm{d}R_{\mathrm{H},\Delta -}}{\mathrm{d}t}\bigg|_{t+\frac{\Delta t}{2}}\Delta t \\ &+ \frac{1}{2} \frac{\mathrm{d}^2R_{\mathrm{H},\Delta -}}{\mathrm{d}t^2}\bigg|_{t+\frac{\Delta t}{2}}(\Delta t)^2
\end{split}
\end{equation}
This corresponds to a second order Taylor expansion, in analogy to the first order Taylor expansion in Eq. \ref{eq:Magnetization}.
Given that \( \left|\frac{1}{2} \frac{\mathrm{d}^2R_{\mathrm{H},\Delta -}}{\mathrm{d}t^2}(\Delta t)^2 \right| \ll \left|\frac{1}{2}\frac{\mathrm{d}R_{\mathrm{H},\Delta -}}{\mathrm{d}t}\Delta t\right| \) which is the case if the time step $\Delta t$ is sufficiently smaller than the time scale on which the relaxation happens, we can neglect the second order term, thus:
\begin{equation}
    R_{\mathrm{H},\Delta+}\left(t+\frac{\Delta t}{2}\right) =  - \frac{1}{2}\frac{\mathrm{d}R_{\mathrm{H},\Delta -}}{\mathrm{d}t}\bigg|_{t+\frac{\Delta t}{2}}\Delta t 
\end{equation}
As shown by the agreement of the red and green curve in Fig. \ref{fig:Decay_sampling_points}(b), since the time scale of the relaxation is slow enough compared to the time step $\Delta t$ to assume, that \(\frac{\mathrm{d}R_{\mathrm{H}}}{\mathrm{d}t}=\frac{\mathrm{d}R_{\mathrm{H},\Delta -}}{\mathrm{d}t}\). In summary, the measurements on the Pt/Co/\ce{AlO_x} trilayer confirm the considerations from Sec. \ref{sec:curr_rev}. Slow magnetic relaxation leads to a measurable signal of the delta plus signal of the Hall resistance, as shown in Fig. \ref{fig:Delta_plus_delta_minus_field_sweep_Co_Pt}(b).
\section{Comparison to the lock-in method}
\label{sec:AC_PtCo}
Another standard method for sensitive (magneto-) resistance measurements and  measurement of higher harmonic signals is homodyne detection using a lock-in amplifier. In the following, we discuss whether lock-in based Hall measurements are equally susceptible to slow relaxation effects. Instead of a dc current, we thus consider that an ac current $I_\mathrm{lock-in}$ with a reference frequency $f_\mathrm{ref}$, the angular frequency $\omega = 2\pi f_\mathrm{ref}$, and an amplitude $I_0$ is applied:
\begin{equation}
    I_\mathrm{lock-in}=I_0\cos({2\pi f_\mathrm{ref}t})=I_0\cos({\omega t})
    \label{eq:LockInCurrent}
\end{equation}
The measured ac voltage $V_\mathrm{m}$ can then be demodulated with respect to different frequencies and phase shifts, so that the in-phase components $V_{x}$ and the quadrature components $V_{y}$ of the first harmonic signal $V^{\omega}$ and the second harmonic signal $V^{2\omega}$ can be extracted.
From Eq. \ref{eq:Hall} and Eq. \ref{eq:LockInCurrent}, it follows, that the slow magnetic relaxation manifests as a decreasing or increasing envelope of the measured ac Hall voltage $V_\mathrm{m}$ (depending on sweep direction and Hall sign).
Due to the changing envelope, the peak at $1\omega$ in Fourier/frequency space is expected to broaden, when a magnetic relaxation is present.
\begin{equation}
    \begin{split}
    V_\mathrm{m} &= (A_\mathrm{OHE}\mu_0 H_\mathrm{ext}\\ & +  A_\mathrm{AHE}\mu_0 M(H_\mathrm{ext},t))\frac{I_0}{d}\cos({\omega t})
    \end{split}
    \label{eq:Hall_ac}
\end{equation}
Therefore, after demodulation with $\cos(\omega t)$ for the in-phase first harmonic of the Hall resistance $R_{\mathrm{H},x}^{\omega}=\frac{V_{\mathrm{H},x}^{\omega}}{I_0}$ in both cases the demodulated signal (e.g. first harmonic of the Hall resistance) is dominated by the time-independent part of the Hall signal. For the first harmonic quadrature component $R_{\mathrm{H},y}^{\omega}=\frac{V_{\mathrm{H},y}^{\omega}}{I_0}$ and the in-phase second harmonic component $R_{\mathrm{H},x}^{2\omega}=\frac{V_{\mathrm{H},x}^{2\omega}}{I_0}$ the measured voltage signal is demodulated with $\sin(\omega t)$ and $\cos(2\omega t)$, respectively. Therefore in the absence of slow magnetic relaxation, the first harmonic quadrature component $R_{\mathrm{H},y}^{\omega}$ and second harmonic in-phase component $R_{\mathrm{H},x}^{2\omega}$ vanish. The residual signals in these components that occur when the magnetic system relaxes is similar to the delta plus signal of the Hall resistance in the dc current-reversal method. In addition, irrespective of whether magnetization relaxation is present, the second harmonic signal can also have contributions from thermoelectric effects, in analogy to the discussion for  the dc-method.\\
\begin{figure}
	\centering
	\includegraphics{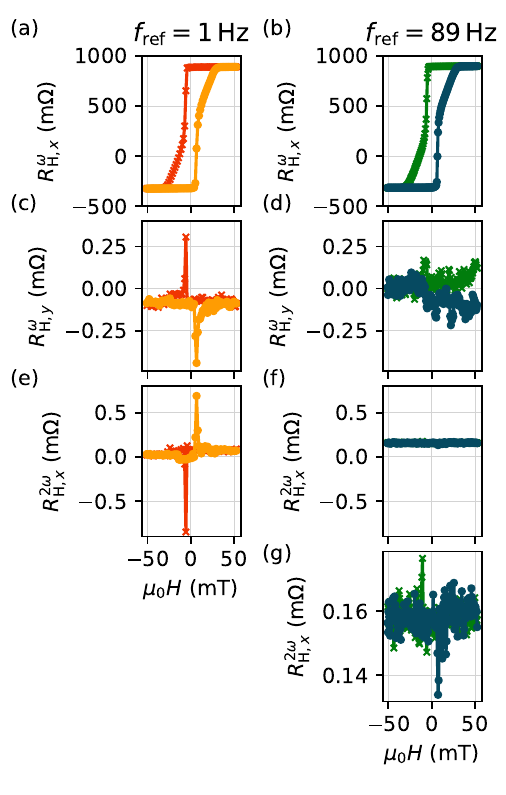}
	\caption{\textbf{ Hall resistance of Pt/Co/AlO$_x$ recorded using the lock-in scheme:}\\
	(a) In-phase component of the first harmonic of the Hall resistance, recorded with a reference frequency of \unit[1]{Hz}. \\(b) In-phase component of the first harmonic of the Hall resistance, recorded with a reference frequency of \unit[89]{Hz}. \\(c) Quadrature component of the first harmonic of the Hall resistance, recorded with a reference frequency of \unit[1]{Hz}. \\(d) Quadrature component of the first harmonic of the Hall resistance, recorded with a reference frequency of \unit[89]{Hz}. \\(e) In-phase component of the second harmonic of the Hall resistance, recorded with a reference frequency of \unit[1]{Hz}. \\(f) In-phase component of the second harmonic of the Hall resistance, recorded with a reference frequency of \unit[89]{Hz}. \\(g) Zoomed plot of the in-phase component of the second harmonic of the Hall resistance, recorded with $f_\mathrm{ref}= \unit[89]{Hz}$. \\The data was recorded at room temperature.
	\label{fig:LockIn_Co_Pt}}
\end{figure}
% 20231031 Pt/Co/AlOx nk220210c2 cross2 RT 1V
To verify these considerations, we performed lock-in measurements on the Pt/Co/AlO$_x$ Hall crosses. In Fig. \ref{fig:LockIn_Co_Pt}(a),(c),(e) we show measurements at a slow reference frequency $f_\mathrm{ref}=\unit[1]{Hz}$. However, in many lock-in measurements higher reference frequencies are used to shift the detection window away from $\frac{1}{f}$ noise, which is large at low $f$ \cite{Keshner1982,Radeka1969}. We therefore also show measurements at a more commonly used reference frequency of \unit[89]{Hz} in Fig. \ref{fig:LockIn_Co_Pt}(b),(d),(f),(g), using a phase correction of $\theta_\mathrm{cor}=\unit[-0.019]{rad}$, to compensate for reactive effects in the circuit. The time constant for the lock-in measurements at $f_\mathrm{ref}=\unit[1]{Hz}$ was chosen as \unit[3]{s}, while the time constant for the measurement at $f_\mathrm{ref}=\unit[89]{Hz}$ was chosen as \unit[1]{s}. For both reference frequencies, a clear Hall signal is present in the in-phase component of the first harmonic of the Hall resistance (see Fig. \ref{fig:LockIn_Co_Pt}(a),(b)). For $f_\mathrm{ref}=\unit[89]{Hz}$, we see no signature of the slow relaxation in the quadrature component of the first harmonic of the Hall resistance (Fig \ref{fig:LockIn_Co_Pt}(d)). Moreover, also the in-phase component of the second harmonic of Hall the resistance (Fig \ref{fig:LockIn_Co_Pt}(f) and (g)) apparently is not affected by relaxation, but rather dominated by leakage of the first harmonic component into the second harmonic component by insufficient filtering or harmonic distortion. For $f_\mathrm{ref}=\unit[1]{Hz}$, small but clear signatures of the slow relaxation in both the quadrature component of the first harmonic of the Hall resistance (Fig \ref{fig:LockIn_Co_Pt}(c)) and the in-phase component of the second harmonic of the Hall resistance (Fig \ref{fig:LockIn_Co_Pt}(e)) appear, as expected. Hence, the lock-in method is rather robust against measurement artifacts
stemming from slow magnetic relaxation effects, at least for reference frequencies $f_\mathrm{ref}\gtrapprox\unit[100]{Hz}$, which are much faster than the typical relaxation rates \cite{Roschewsky2019,Avci2014}.
\section{Magnetic relaxation in M\lowercase{n}A\lowercase{l} thin films}
\label{sec:DC_MnAl}
\begin{figure}
	\centering
	\includegraphics{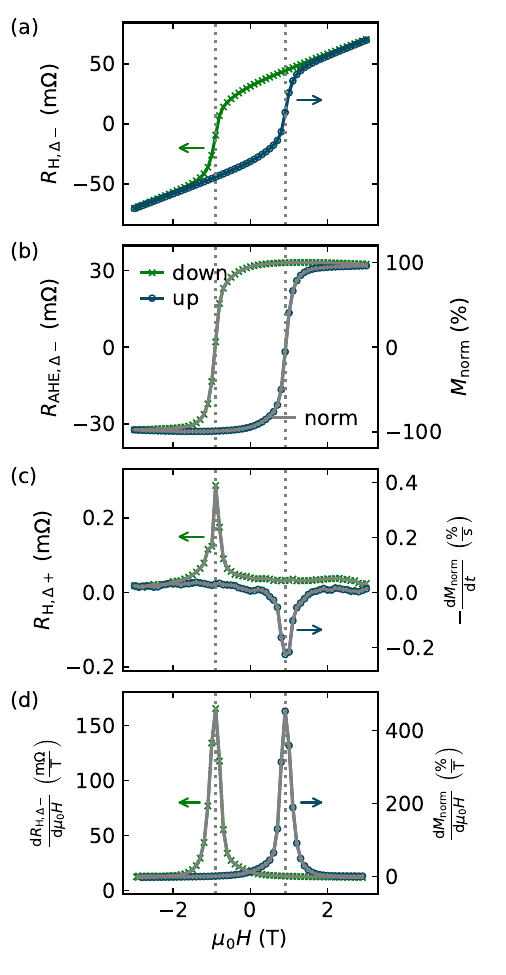}
	\caption{\textbf{Magnetic relaxation in a MnAl thin film detected via the delta method:} \\(a) Delta minus signal of the Hall resistance. \\(b) Left axis: delta minus signal of the anomalous Hall resistance obtained by substracting the OHE from the data in (a), right axis: normalized magnetization calculated from the anomalous Hall resistance, using Eq. \ref{eq:M_norm}. \\(c) Left axis: delta plus signal of the Hall resistance, right axis: negative of the (normalized) relaxation parameter \(\frac{\mathrm{d}M_\mathrm{norm}}{\mathrm{d}t}\) calculated from the delta plus signal of the Hall resistance via Eq. \ref{eq:time_deriv}. \\(d) Derivative of the delta minus signal of the Hall resistance with respect to the external field $\mu_0 H$ at $T=\unit[300]{K}$, recorded with a time step $\Delta t=\unit[4.5]{s}$ and $I=\unit[5]{mA}$. The field sweep directions are indicated by arrows and different colors. \label{fig:Delta_plus_delta_minus_field_sweep_MnAl}}
\end{figure}
% 20231005 MnAl ds7210930a 300K 5mA 
We now use the current-reversal method to gauge the presence of magnetic relaxation in a material of current interest, $\tau$-MnAl. This ferromagnetic compound features a high uniaxial anisotropy leading to a PMA behaviour, which makes it an interesting candidate for spintronic applications \cite{Coey2014,Scheffler2023}. Slow magnetic relaxation processes have recently been observed in ferromagnetic $\tau$-Mn-Al-C alloys \cite{Pasco2021}. We here focus on a  \unit[78]{nm} thick MnAl film sputter-deposited onto a \unit[19]{nm} thick Cr buffer layer on a MgO(001) substrate. Details of the growth parameters of the film are described by Scheffler \textit{et al.} \cite{Scheffler2023}. The film is patterned into a Hall bar by optical lithography and wet etching. The Hall response is measured in a superconducting magnet cryostat with variable temperature insert, since the coercive field of \ce{MnAl} is much higher than the coercive field of the Pt/Co/AlO$_x$ film. Again, the current was applied using a Keithley 2450 sourcemeter and the transversal voltage measured with a Keithley 2182 nanovoltmeter.
In Fig. \ref{fig:Delta_plus_delta_minus_field_sweep_MnAl}(a) we show the Hall resistance (recorded using the delta method) in dependence of the external magnetic field. In Fig. \ref{fig:Delta_plus_delta_minus_field_sweep_MnAl}(b) (left axis) the OHE contribution is removed from the Hall measurements, as described in Sec. \ref{sec:curr_rev}, such that only the delta minus signal of the anomalous Hall resistance remains. On the right axis the normalized magnetization is depicted, which is calculated using Eq. \ref{eq:M_norm}. The delta plus signal of the Hall resistance is shown in the left axis of Fig. \ref{fig:Delta_plus_delta_minus_field_sweep_MnAl}(c). This signal shows clear peaks when the magnetization reorients, indicating the presence of slow relaxation of the magnetization. As can be seen on the left axis of Fig. \ref{fig:Delta_plus_delta_minus_field_sweep_MnAl}(d) the maxima of the delta plus signal again coincide with the maximum change of Hall resistance and therefore maximum change of magnetization (i.e., maximal susceptibility) in the sample. In the right axis of Fig. \ref{fig:Delta_plus_delta_minus_field_sweep_MnAl} the negative of the normalized relaxation parameter \(\frac{\mathrm{d}M_\mathrm{norm}}{\mathrm{d}t}\) is depicted, using Eq. \ref{eq:time_deriv}. Here the maximum of the normalized relaxation parameter is approx. \unitfrac[0.4]{\%}{s}. This normalized relaxation parameter is an order of magnitude smaller compared to the \unitfrac[8]{\%}{s} observed in the Pt/Co/\ce{AlO_x} trilayer. We therefore expect, that the time scale of relaxation in the MnAl sample is higher (i.e., slower relaxation) compared to the Pt/Co/\ce{AlO_x} sample. Using the approximation from Eq. \ref{eq:dHdt}, the normalized relaxation parameter \(\frac{\mathrm{d}M_\mathrm{norm}}{\mathrm{d}t}\) (see right axis Fig. \ref{fig:Delta_plus_delta_minus_field_sweep_MnAl}(c)), and the normalized susceptibility \(\frac{\mathrm{d}M_\mathrm{norm}}{\mathrm{d}\mu_0 H}\) (see right axis of Fig. \ref{fig:Delta_plus_delta_minus_field_sweep_MnAl}(d)) we can estimate the sensitivity of the measured coercive field with regard to small changes of the field sweep rate or measurement delays as approx. \unitfrac[0.0009]{T}{s}, which is negligible small compared to the coercive field $H_\mathrm{c}\approx\unit[0.9]{T}$. Therefore we expect no noticeable influence of (reasonable) changes of the field sweep rate on the measurement of the coercive field.
\begin{figure}
	\centering
	\includegraphics{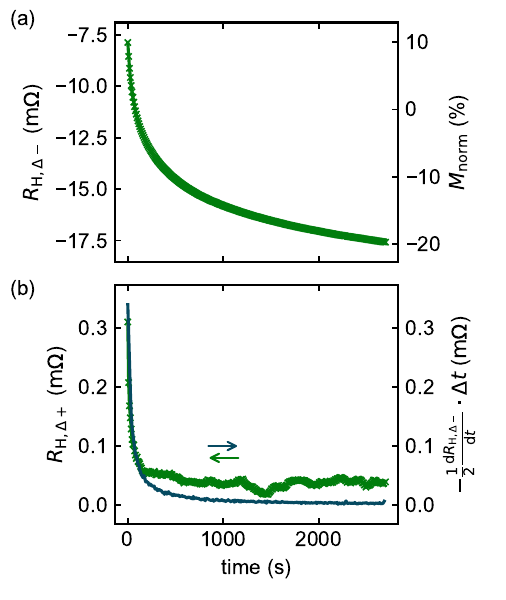}
	\caption{\textbf{Time-dependent evolution of the Hall resistance in a MnAl film at $\mu_0 H = \unit[-0.9]{T}$:}\\
	(a) Left axis: measured time trace of the delta minus signal of the Hall resistance $R_{\mathrm{H},\Delta-}$ , right axis: time trace of the normalized magnetization calculated from $R_{\mathrm{H},\Delta-}$ using Eq. \ref{eq:M_norm}. \\(b) Left axis: measured time trace of the delta plus signal of the Hall resistance (green), right axis: derivative $-\frac{1}{2}\frac{\mathrm{d}R_{\mathrm{H},\Delta-}}{\mathrm{d}t}$ (blue) calculated via Eq. \ref{eq:R_trans_plus}.}\label{fig:decay_MnAl}
\end{figure}
To confirm this assumption, we measured a time trace of the transversal voltage. Like in Sec. \ref{sec:DC_PtCo} for this measurement, first the magnetization of the sample is saturated at a high (positive) magnetic field after which the external field is ramped to a negative field value (i.e., $\unit[-0.9]{T}$) and then kept constant. Once the field is stable, the voltage is recorded over time using the delta method. In Fig. \ref{fig:decay_MnAl}(a) we show the delta minus signal of the Hall resistance over time for this measurement. The normalized magnetization values corresponding to this Hall resistances are depicted in the right axis. Fig. \ref{fig:decay_MnAl}(b) shows the delta plus signal of the Hall resistance on the left axis (green curve) and the correspondence of this values to the term \(-\frac{1}{2}\frac{\mathrm{d}R_{\mathrm{H},\Delta-}}{\mathrm{d}t}\Delta t\) (see Eq. \ref{eq:R_trans_plus}) on the right axis (blue curve). Note that the curves only match until a time of approx. \unit[150]{s}. After this time, the measured curve is dominated by noise, since the theoretical calculated voltage is beneath the noise floor of the experimental setup and relaxation-independent thermoelectric voltages from the setup. However, as evident from the measurement curve, the relaxation in the MnAl thin film happens on a much longer time scale (cf. time axis in Fig. \ref{fig:decay_MnAl}) as compared to the relaxation in the Pt/Co/\ce{AlO_x} trilayer (see time axis in Fig. \ref{fig:Decay_sampling_points})  This corroborates the notion that the current-reversal method allows to detect the (normalized) relaxation parameter and subtle changes in slow magnetic relaxation. In particular, in  field sweep rate dependent measurements on MnAl, such subtle changes could easily be overlooked, highlighting a big advantage of the current-reversal method.\section{Summary and outlook}
\label{sec:summ}
In conclusion, we presented a scheme for the detection of slow magnetic relaxation effects based on the measurement of the AHE response via a standard dc current-reversal method. This method can easily be applied to standard Hall measurements and is very sensitive to even subtle relaxation processes. As such, it is an ideal technique for fast screening, providing first information about the time scale and amplitude of slow relaxation processes quantify $\frac{\mathrm{d}M_\mathrm{norm}}{\mathrm{d}t}$, and identify field ranges for further investigation of the relaxation processes in magnetic films showing an AHE. Our analysis reveals that the method is ideally suited to study relaxation processes on the time scale of a few seconds up to several hundred seconds. For such relaxation parameters, conventional lock-in based magnetotransport measurements are much less sensitive to relaxation effects. In other words, the dc delta method has the big advantage, that it works at very low frequencies, for which lock-in schemes are overwhelmed by $\frac{1}{f}$ noise.

\section*{Acknowledgments}
This work was funded by the Deutsche Forschungsgemeinschaft (DFG, German Research Foundation)  project IDs 445976410 and 490730630, and via the SFB 1432 - Project-ID 425217212. This work was supported by the French national research agency (ANR) project ID ANR-20-CE92-0049-01. DK acknowledges the Czech Academy of Sciences (project no. LQ100102201). HR was supported by the Grant Agency of the Czech Republic Grant No. 22-17899K and the Dioscuri Program LV23025.
\bibliography{main.bbl}

\end{document}